\newcommand{\be}{\beta}

\newcommand{\fr}{\frac}
\newcommand{\Pl}{\partial}

\newcommand{\ts}{\textstyle}
\newcommand{\bee}{\begin{equation}}
\newcommand{\ene}{\end{equation}}
\newcommand{\beea}{\begin{eqnarray}}
\newcommand{\enea}{\end{eqnarray}}

\newcommand{\fder}[2]{\frac{{\ts d \/ #1}}{{\ts d\/ #2}}}
\newcommand{\fpar}[2]{\frac{{\ts \Pl \/ #1}}{{\ts \Pl \/ #2}}}
\newcommand{\nder}[3]{\frac{{\ts d^{#1} \/ #2}}{{\ts d \/ #3^{#1}}}}

\documentclass[pre,twocolumn]{revtex4}
\usepackage{graphicx}
\usepackage{amsmath,amsfonts,amsthm,eucal,amssymb}
\usepackage{bm}
\begin{document}
 \title{Wave Breaking limit in Arbitrary Mass Ratio Warm Plasmas}
\author{ Ashish Adak$^{1}$\footnote{ashish$_{-}$adak@yahoo.com}, Nidhi Rathee$^{1,2}$ and Sudip Sengupta$^{1,2}$}
    \affiliation{$^{1}$Institute for Plasma Research, Bhat, Gandhinagar-382428, India.\\$^{2}$Homi Bhabha National Institute, Training School Complex, Mumbai 400094, India}
    %
\begin{abstract}
The maximum sustainable amplitude, so-called wave breaking limit, of a nonlinear plasma wave in  arbitrary mass ratio warm plasmas is obtained in the non-relativistic regime. Using the method of Sagdeev potential a general wave breaking formula is derived by taking into account the dynamics of both the species 
having finite temperature. 
%
It is found, that the maximum amplitude of the plasma wave decreases monotonically with the increase in temperature and mildly increases with increase in  mass ratio. 
\end{abstract}

 \maketitle

Studies on wave breaking is a topic of fundamental interest in plasmas due to its various applications such as plasma heating \cite{ah74,pk74}, plasma based particle acceleration schemes \cite{cj84,am95,ja75,rb07}, etc. The wave breaking \cite{daw59,rc68,rc72,vehn} limit of a nonlinear plasma oscillation and/or wave decides its maximum sustainable amplitude beyond which the coherent nature of the wave is destroyed. At the wave breaking point, the velocity of the plasma fluid element at the crest of the wave exceeds the phase velocity of the wave.
%
In certain situations, a wave breaks at a lower amplitude than its usual breaking limit because of a novel phenomena called 'phase mixing' \cite{tk83,ss99,ss09,cm12}. Phase mixing is an important physical process through which an oscillation or/and wave undergoes a gradual loss of phase coherence, thus leading to breaking of the wave at a finite time, even if the amplitude of the wave is well below the breaking amplitude \cite{verma}. Phase mixing occurs when the frequency of the wave becomes space-dependent and this may occur through various  nonlinear processes like inhomogeneity \cite{ref1}, relativistic mass variation \cite{ref2}, etc.

The maximum amplitude or the wave breaking amplitude of an electron plasma wave was first introduced by Akhiezer and Polovin (AP) \cite{ap56} in a cold relativistic plasma where the massive ions were assumed to provide a fixed charge neutralizing background. Although, they never use the term ``wave breaking'' in their article, the derived wave breaking limit was $E_{wb} =(mc\omega_{pe}/e)\sqrt{2(\gamma-1)}$, where $\gamma=[1-(v_{ph}/c)^2]^{-1/2}$, $v_{ph}$ is the phase velocity of the wave, $\omega_{pe}$ is the electron plasma frequency and $m$ is the mass of an electron.  Later, Dawson\cite{daw59} derived the wave breaking amplitude $E_{wb}$ in the nonrelativistic limit using a Lagrangian description which gives $E_{wb}=mv_{ph}\omega_{pe}/e$. Again in the nonrelativistic regime, thermal effects were included in the study of wave breaking by Coffey \cite{cof71} using a 1D waterbag model for electrons. The maximum sustainable amplitude derived by Coffey was $E_{wb}=(mv_{ph}\omega_{pe}/e)\left(1 - \fr{1}{3} \beta - \fr{8}{3}\beta^{1/4} + 2\beta^{1/2}\right)^{1/2}$, where $\beta=3T/mv_{ph}^2$, $T$ is the temperature in the energy unit. It was found that the inclusion of the temperature through the plasma pressure reduces the wave breaking amplitude. Katsouleas and Mori \cite{mk88} investigated the wave breaking limit in the relativistic regime and obtained the wave breaking limit as $E_{wb}=(mc\omega_{pe}/e)\beta^{-1/4}(\mbox{ln}2\gamma^{1/2}\beta^{1/4})^{1/2}.$  
%
Thus the subject of wave breaking of large amplitude plasma waves with immobile ions has been thoroughly investigated with both cold and warm electrons, spanning the entire domain, starting from non-relativistic to relativistic regime. Khachatryan \cite{khac} extended these studies on wave breaking by including ion motion in a cold relativistic electron-ion plasma. It was reported that, with the increase of electron to ion mass ratio, the wave breaking amplitude also increases. 
As a corollary, it was shown that the wave breaking limit in non-relativistic cold pair-ion plasmas 
(e.g., electron-positron plasmas/pair-ion plasmas) is higher than electron-ion plasmas ( $E_{wb}=1.08(mv_{ph}\omega_{pe}/e)$ ). However, to the best of our knowledge wave breaking studies with both warm electrons and warm ions have never been attempted.
%
In this letter, we study finite temperature effects on wave breaking in warm
unmagnetized arbitrary mass ratio plasmas, in the non-relativistic regime.

To obtain the general wave breaking amplitude, we have
considered an unmagnetized, homogeneous warm nonrelativistic
plasma with two species having equal and opposite charges.
%
%
In one space-dimension, the basic equations that govern the propagation of nonlinear electrostatic waves in warm plasmas are the continuity equations, momentum equations for positively(ions or positrons) and negatively(electrons or negative ions ) charged particles, and the Poisson's equation, which are respectively written as follows:
\bee
    \fpar{n_\pm}{t}+\fpar{}{x}(n_\pm u_\pm)=0,
    \label{con}
    \ene
\bee
    \left(\fpar{}{t}+u_\pm\fpar{}{x}\right)u_\pm=\pm\fr{e E}{m_\pm}
    -\fr{1}{m_\pm n_\pm}\fpar{P_\pm}{x},
    \label{mom}
    \ene
and
\bee
    \fpar{E}{x}=4 \pi e \left(n_+ - n_-\right),
    \label{poi}
    \ene
where $n_\pm, u_\pm, P_\pm$ and $m_\pm$ are densities, velocities,
partial pressures and masses of positively and negatively charged particles
respectively; and  
$E$ is the electric field. The plasma is overall quasi-neutral, and for
singly charged particles we assume quasineutrality
condition as $n_{+0}=n_{-0}=n_0 (\mbox{say})$, where $n_{+0}$ and
$n_{-0}$ are respectively the equilibrium number density of positively and
negatively charged particles. Using the adiabatic equation of state for both
positively and negatively charged particles with $\gamma = 3$ (the ratio of specific heats,
for a 1-D system), the equations of motion may be written as
\bee
    \left(\fpar{}{t}+u_\pm\fpar{}{x}\right)u_\pm=\pm\fr{e E}{m_\pm}
    -\fr{v_{th \pm}^{2}}{2 n_{0}^{2}}\fpar{n_\pm^2}{x},
    \label{mom_mod}
    \ene
%
%
where $v_{th\pm}=\sqrt{3 T_\pm/m_\pm}$ is the thermal speed of each species
with $T_+$ and $T_-$ as the equilibrium temperatures of the positively and negatively charged particles respectively. The governing equations (\ref{con}), (\ref{poi}) and (\ref{mom_mod}) are exact within the framework of waterbag model, {\it i.e.} these equations may be directly derived by taking moments of the Vlasov equation assuming waterbag distribution for the warm species. For such a distribution the heat flux turns out to be identically zero so that there is closure to the hierarchy of moments of the  Vlasov equation \cite{rc72}. 

For traveling wave solution of equations (\ref{con}), (\ref{poi})
and (\ref{mom_mod}), it is convenient to move into a wave frame
with the variable transformation $\psi=k_p\left(x - v_{ph}
t\right)$, $k_p=\omega_p/v_{ph}$, where plasma frequency
$\omega_p=\sqrt{4\pi n_0 e^2/m_-}$ and $v_{ph}$ is the phase
velocity of the wave. With this transformation the equations
(\ref{con}), (\ref{mom_mod}) and (\ref{poi}) transforms, respectively into the following ordinary differential equations:
\bee
    \fder{}{\psi}\left[\hat{n}_\pm \left(1-\hat{u}_\pm\right)\right]=0,
    \label{con1}
    \ene

\bee
    \fder{}{\psi}\left[\hat{u}_{\pm}^{2} - 2\hat{u}_\pm \right]=\pm 2 \mu_\pm
    \hat{E} - \beta_\pm \fder{}{\psi}\left( \hat{n}_{\pm}^{2} \right),
    \label{mom1}
    \ene
and
\bee
    \fder{\hat{E}}{\psi}=\hat{n}_+ - \hat{n}_-\,
    \label{poi1}
    \ene
where $\hat{n}_\pm, \hat{u}_\pm$ and $\hat{E}$ are the normalized
densities, velocities and electric field, normalized by
equilibrium density $n_0$, phase velocity $v_{ph}$ and $m_-
\omega_p v_{ph}/e$, respectively. The other plasma parameters are
mass ratio $\mu_\pm=m_-/m_\pm$ (here $\mu_-=1$ and
$\mu_+=m_-/m_+=\mu \leq 1$ (say)), and $\beta_\pm = v_{th\pm}^2/v_{ph}^2$.
The normalized electric field can be expressed as $\hat{E}=-d\hat{\phi}/d\psi$,
$\hat{\phi}=e\phi/m_-v_{ph}^2$ is the normalized electric
potential. Using this expression for normalized electric field, 
Eqs. (\ref{con1}) and (\ref{mom1}) respectively reduces to 
\bee
    \hat{n}_\pm =\fr{1}{\left(1-\hat{u}_\pm\right)}
    \label{con2}
    \ene
\bee
    \left(1-\hat{u}_\pm\right)^2=\fr{\phi_\pm + \sqrt{\phi_{\pm}^{2} - 4     \beta_\pm}}{2},
    \label{mom2}
    \ene
where we have chosen $\hat{n}_\pm=1$, and $\hat{\phi}=0$
for $\hat{u}_\pm=0$. In (Eq. (\ref{mom2})), we have
set $\phi_-= 1 + \beta_- +2 \hat{\phi}$ and $\phi_+= (1 +\beta_+) +
\mu (1+ \beta_-) - \mu \phi_-$. Using Eqs. (\ref{con2})
and (\ref{mom2}), from Eq. (\ref{poi1}) we obtain the following
second order ordinary differential equation:
\beea
    \fr{1}{2}\nder{2}{\phi_-}{\psi} + \fr{\sqrt{2}}{\left(\phi_+ +
    \sqrt{\phi_+^2-4\beta_+}\right)^{1/2}}\nonumber \\
    -\fr{\sqrt{2}}{\left(\phi_- + \sqrt{\phi_-^2-4\beta_-}\right)^{1/2}}= 0
    \label{energy}
    \enea
Here we have used $\hat{E}=-(1/2)(d\phi_-/d\psi)$. The above equation may be further rewritten as
%
%
\begin{figure}
\centering {\includegraphics[width=3in,height=4in]{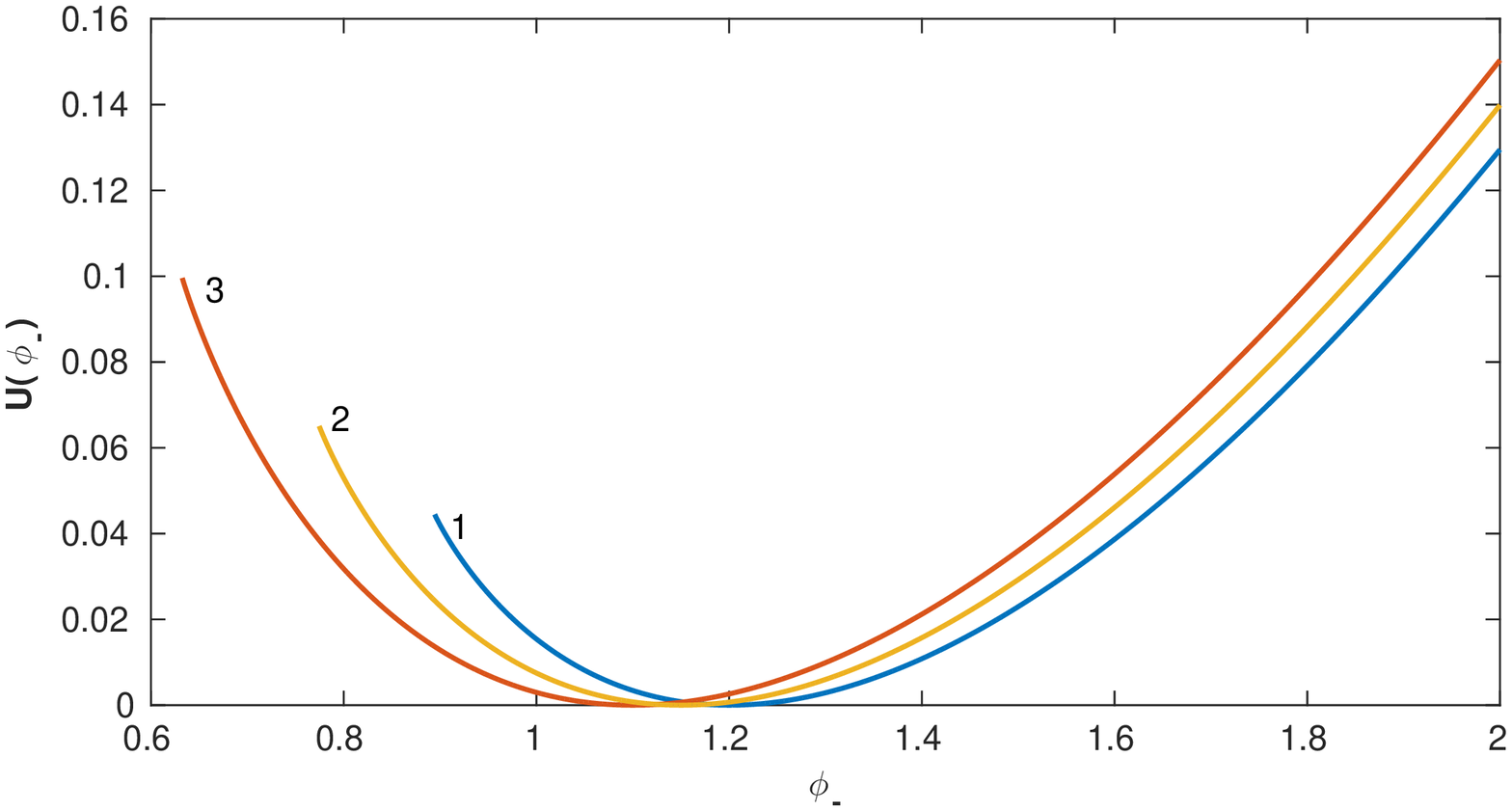}}
             \caption{Potential $U(\phi_-)$ vs. $\phi_-$ for the plasma
             parameter $\mu=1/1836$ and 1 $\rightarrow \beta_+(=0.1) < \beta_-(=0.2)$,
             2 $\rightarrow \beta_+ = \beta_-= 0.15$ , 3 $\rightarrow \beta_+(=0.2) > \beta_-(=0.1)$.}
             \label{x1}
             \end{figure}
\begin{figure}
\centering
{\includegraphics[width=3in,height=4in]{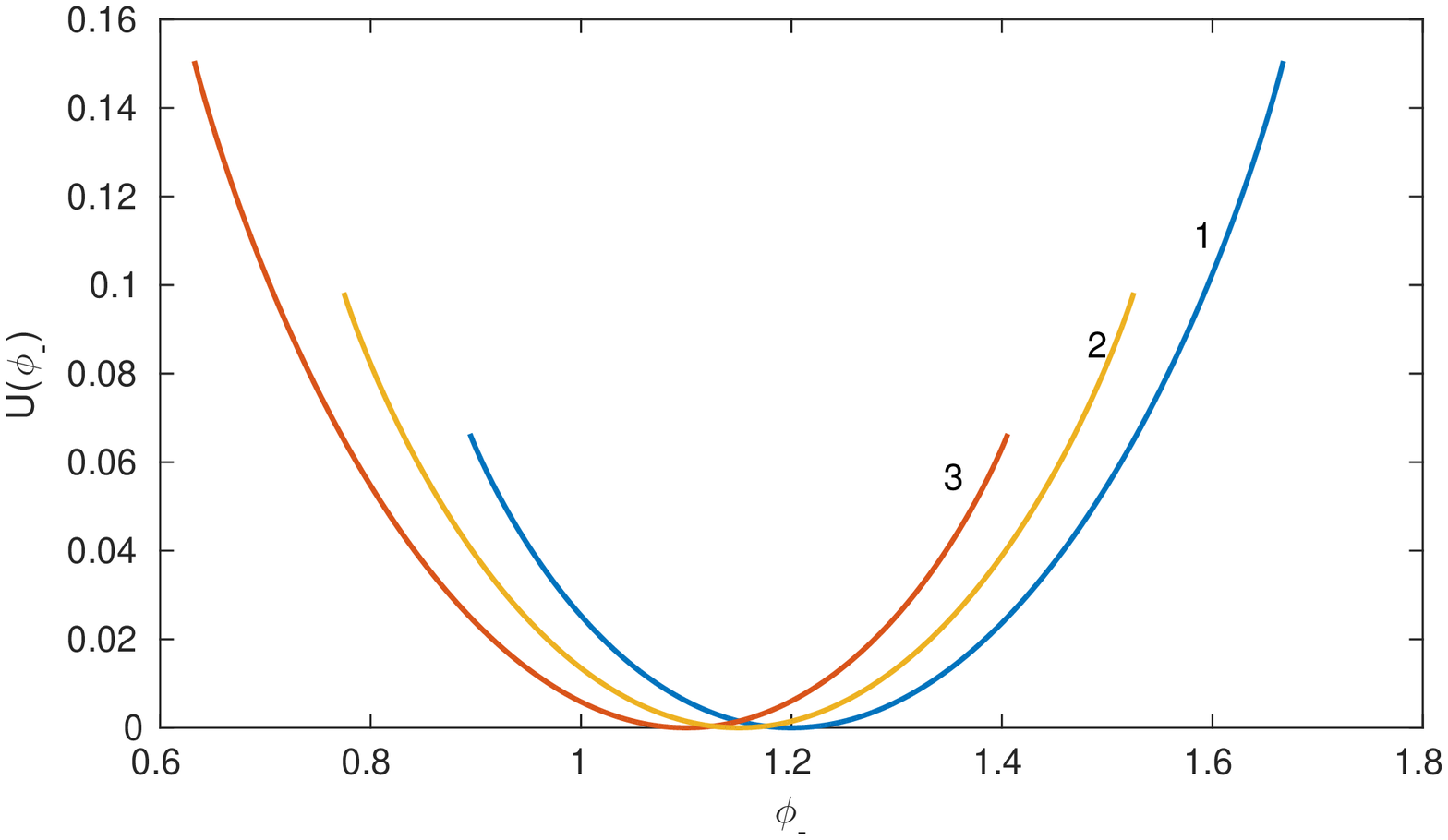}}
             \caption{Potential $U(\phi_-)$ vs. $\phi_-$ for the plasma
             parameter $\mu=1.0$ and 1 $\rightarrow \beta_+(=0.1) < \beta_-(=0.2)$,
             2 $\rightarrow \beta_+=\beta_-=0.15$ , 3 $\rightarrow \beta_+(=0.2) > \beta_-(=0.1)$.}
             \label{x2}
             \end{figure}

\bee
    \fr{1}{2}\nder{2}{\phi_-}{\psi} + \fder{U}{\phi_-}= 0,
    \label{energy1}
    \ene
where $U(\phi_-)$ is the Sagdeev potential, which is given by
%
%
\beea
    U(\phi_-) = \sqrt{2}\left[\left(\sqrt{2}-\xi_2\right)+\fr{\sqrt{2}\beta_-}{3}\left(1-\fr{2\sqrt{2}}{\xi_2^3}\right)\right]\nonumber\\
    +\fr{\sqrt{2}}{\mu}\left[\left(\sqrt{2}-\xi_1\right)+\fr{\sqrt{2}\beta_+}{3}\left(1-\fr{2\sqrt{2}}{\xi_1^3}\right)\right]
    \label{sagdeev}
    \enea
Here $\xi_{1,2}(\phi_-)=\left(\phi_\pm + \sqrt{\phi_\pm^2-4\beta_\pm}\right)^{1/2}$
and $U(\phi_-)$ is chosen to be equal to zero at $\hat{\phi}=0$, i.e., at $\phi_-=1+\beta_-$.
It is clear from expression (\ref{sagdeev}) that for real values of $U(\phi_-)$ the range of $\phi_-$ is
$\phi_1 \leq \phi_-\leq \phi_2 $ where $\phi_1 = 2 \beta_-^{1/2}$ and $\phi_2 = 1+\beta_- + (1/\mu)( 1 - \beta_+^{1/2} )^2$. Within this range, periodic solutions to (\ref{energy1}) may exist. 
%
%
The first integral of equation (\ref{energy1}) gives
\bee
    \hat{E}^2 + U = U_{max},
    \label{energy2}
    \ene
where $U_{max}$ is the integration constant indicating the total
energy of a fictitious particle obeying the differential equation
(\ref{energy1}) and is the maximum permissible value of
$U(\phi_-)$.
%
%
Therefore, the maximum achievable electric field, i.e.,the wave
breaking amplitude is $\hat{E}_{wb}=\sqrt{U_{max}}$. This is
the expression for wave breaking amplitude in arbitrary mass ratio warm plasmas. It is clear from Eq.(\ref{sagdeev}) that the magnitude of wave breaking amplitude depends on the parameters $\mu$ and $\beta_{\pm}$. 

Below we explore the wave breaking amplitude in arbitrary mass ratio warm plasmas for some typical values of $\mu$; we have chosen the value of $\mu=1/1836$
and $\mu=1.0$, the first value of $\mu$ is for warm electron-ion
plasmas and later one is for warm pair ion plasmas. 
From Fig.
\ref{x1}, it is clear that for $\mu=1/1836$, periodic solutions are possible upto $U_{max}$
calculated at $\phi_-=\phi_1$ for $\beta_+ > = < \beta_-$, which
implies that for $\mu = 1/1836$ and $\beta_+ > = < \beta_-$
\bee
    \hat{E}_{wb} =\sqrt{ U(\phi_1)},
    \label{wb}
    \ene
where
\beea
    U(\phi_1) =2\left[\fr{1}{\mu} + 1 + \fr{\beta_+}{3\mu}+\fr{\beta_-}{3} -\fr{4}{3}\beta_-^{1/4}\right] \nonumber\\
    -\fr{3\sqrt{2}\left(\phi_{*} + \sqrt{\phi_{*}^2-4\beta_+}\right)^{2} + 4\sqrt{2}\beta_+}{3\mu\left(\phi_{*} +
    \sqrt{\phi_{*}^2-4\beta_+}\right)^{3/2}}\nonumber
    \enea
Here we define\[\phi_{*}=\phi_+|_{\phi_-=\phi_1} = (1+\beta_+) +
\mu (1 - \beta_-^{1/2})^2 .\] Therefore, for $\mu=1/1836$, wave
breaking amplitude does not depend on the relative values of  $\be_-$
and $\be_+$. However, for $\mu=1$, wave breaking amplitude does
depend on the relative values of $\be_-$ and $\be_+$ (see Fig. \ref{x2}). There are two
different wave breaking limit for $\mu=1.0$ depending on whether
$\be_->\be_+$, or $\be_-<\be_+$, i.e., equivalently, $T_-> T_+$ or
$T_-< T_+$. Fig. \ref{x2} shows that, the wave breaking amplitude
$\hat{E}_{wb} = \sqrt{ U(\phi_1)}$ for 
$\be_->\be_+$
i.e., equation (\ref{wb}) with $\mu=1.0$.
In contrast, for $\be_-<\be_+$, Fig. \ref{x2} shows that, the wave breaking amplitude $\hat{E}_{wb} = \sqrt{ U(\phi_2)}$ 
where, $\phi_2 = 1+\beta_- + (1 - \beta_+^{1/2} )^2$. 
Therefore, the wave breaking amplitude of the plasma
wave for $\mu=1.0$ and $\be_-<\be_+$ is
\bee
    \hat{E}_{wb} =\sqrt{ U(\phi_2)},
    \label{wb1}
    \ene
where
\beea
    U(\phi_2) =2\left[2+ \fr{\beta_+}{3}+\fr{\beta_-}{3} -\fr{4}{3}\beta_+^{1/4}\right] \nonumber\\
    -\fr{3\sqrt{2}\left(\phi_2 + \sqrt{{\phi_2}^2-4\beta_-}\right)^{2} + 4\sqrt{2}\beta_-}{3\left(\phi_2 +
    \sqrt{\phi_2^2-4\beta_-}\right)^{3/2}}\nonumber
    \enea
Here we note that wave breaking amplitudes Eq. (\ref{wb}) and
Eq. (\ref{wb1}) for $\mu = 1.0$ and $\be_- =
\be_+$ are same. 

For large amplitude electron plasma waves in a warm plasma with immobile ions, from Eq.(\ref{wb}), with $\mu\rightarrow0$, $\beta_+=0$
and $\beta_-=\beta$, we recover Coffey's limit \cite{cof71} as 
\bee
    \hat{E}_{wb} = \left(1 - \fr{1}{3} \beta - \fr{8}{3}\beta^{1/4} + 2 \beta^{1/2}
    \right)^{1/2},
    \ene
When ion dynamics is included, for large amplitude electron-ion waves in cold plasmas, by setting $\beta_+=0$ and $\beta_-=0$ in Eq. (\ref{wb}) we obtain
\bee
    \hat{E}_{wb} = \left[2\fr{\sqrt{1+\mu}}{\mu}\left( \sqrt{1+\mu}-1\right) \right]^{1/2}
    \label{emax_kha}
    \ene
This is the non-relativistic version of the wave breaking limit derived by Khachatryan \cite{khac}. By  setting $\mu=1$ in the above equation, we obtain $\hat{E}_{wb}=1.08$ for large amplitude waves in pair-ion plasmas. 

\begin{figure}[ht]
\centering {\includegraphics[width=3in,height=1.5in]{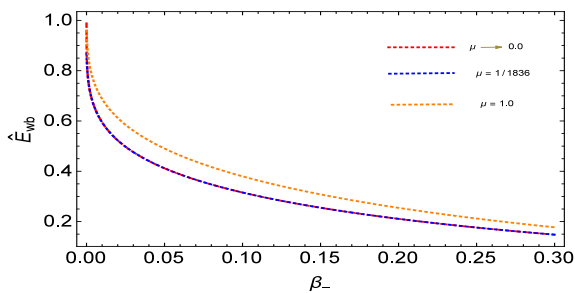}}
             \caption{Maximum sustainable amplitude $\hat{E}_{wb}$  vs. $\beta_-$. $\beta_+=0$ for  $\mu\rightarrow0.0$
             (Coffey's $\hat{E}_{wb}$); fixed $\beta_+=0.0001$ for
             $\mu=1/1836$ ($\hat{E}_{wb}$ for electron-ion plasmas); fixed $\beta_+=0.0001$ for $\mu=1.0$ ($\hat{E}_{wb}$ for pair-ion plasmas)  }
             \label{x3}
             \end{figure}
\begin{figure}[ht]
\centering
{\includegraphics[width=3in,height=1.5in]{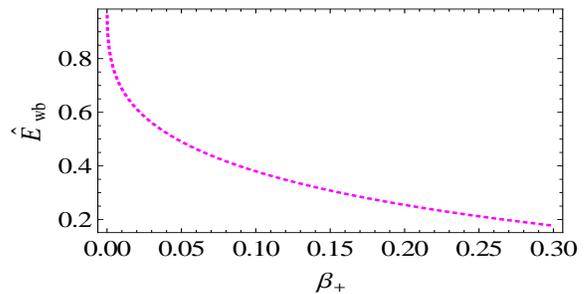}}
             \caption{Maximum sustainable amplitude $\hat{E}_{wb}$  vs. $\beta_+$ for a fixed $\beta_-=0.0001$ and $\mu=1.0$.}
             \label{x4}
             \end{figure}

In Fig. \ref{x3} and \ref{x4}, we have plotted the normalized wave breaking electric field $(\hat{E}_{wb})$ vs. normalized thermal speed $\be_- (\be_+)$ respectively given by Eq. (\ref{wb}) and Eq.  (\ref{wb1}), for fixed $\be_+ (\be_-)$. In Fig.
\ref{x3}, Coffey's wave breaking limit
$(\mu\rightarrow 0)$ is compared with the wave breaking limit for 
$(\mu=1/1836)$ and the wave breaking limit for pair ion plasmas
$(\mu=1.0)$. In all the cases, $\hat{E}_{wb}$ decreases with the increase of $\be_-$ for a fixed $\beta_+=0.0001$ (for Coffey's case $\be_+=0$). It is also observed that for a fixed temperature ({\it i.e.} $\be_-$), the wave breaking amplitude mildly increases with increasing $\mu$. This feature has also been reported by Khachatryan \cite{khac} for the cold relativistic case. In Fig. \ref{x4}, we have shown the variation
of $\hat{E}_{wb}$ given by Eq. (\ref{wb1}) vs. $\be_+$ for a fixed
$\beta_-=0.0001$.

%

In conclusion, we have derived the general wave breaking amplitude
in nonrelativistic unmagnetized plasmas with finite
temperature for both species and different mass ratios. From the
general wave breaking amplitude, we have recovered the earlier results derived by Coffey \cite{cof71} and Khachatryan \cite{khac}.
%
It is found that
the wave breaking amplitude ($\hat{E}_{wb}$) of a plasma wave in arbitrary mass ratio plasmas depends on the thermal to phase velocity ratio
$(\be_\pm)$ and it decreases monotonically with the increase of $\be_\pm$.

Author A. Adak would like to express his gratitude to the Science
and Engineering Research Board (SERB), Department of Science and
Technology (DST), India, for providing research grant under
National Post-Doctoral Fellowship (the fellowship reference No.
PDF/2017/001750) for this work.

\end{document}